# Specific heat analyses on optical-phonon-derived uniaxial negative thermal expansion system $Tr$Zr$_2$ ($Tr$ = Fe and Co$_{1-x}$Ni$_x$)

Yuto Watanabe[1], Ceren Tayran[2,3]*, Md. Riad Kasem[1], Aichi Yamashita[1], Mehmet Çakmak[4,5], Takayoshi Katase[6], Yoshikazu Mizuguchi[1]*

[1]Department of Physics, Tokyo Metropolitan University, Hachioji, Tokyo 192-0397, Japan
[2]Department of Physics, Faculty of Science, Gazi University, 06500, Ankara, Turkey
[3]DIFFER–Dutch Institute for Fundamental Energy Research, De Zaale 20, 5612 AJ Eindhoven, The Netherlands
[4]Department of Photonics, Faculty of Applied Sciences, Gazi University, 06500, Ankara, Turkey
[5]Photonics Application and Research Center, Gazi University, 06500, Ankara, Turkey
[6]MDX Research Center for Element Strategy, International Research Frontiers Initiative, Tokyo Institute of Technology, Yokohama 226-8501, Japan

Corresponding authors: C. Tayran (c.tayran@gazi.edu.tr), Y. Mizuguchi (mizugu@tmu.ac.jp)

**Abstract**

Recently, huge uniaxial negative thermal expansion (NTE) along a $c$-axis has been observed in transition-metal ($Tr$) zirconides $Tr$Zr$_2$ with a tetragonal CuAl$_2$-type structure. In a recent study on FeZr$_2$ [M. Xu et al., Nat. Commun. 14, 4439 (2023)], the importance of optical phonons to the emergence of the $c$-axis NTE in FeZr$_2$ has been proposed. In this study, the physical properties of $Tr$Zr$_2$ ($Tr$ = Fe and Co$_{1-x}$Ni$_x$) have been studied by specific heat, sound velocity measurements, and theoretical phonon calculations to discuss the importance of optical phonons to the emergence of the $c$-axis NTE in CoZr$_2$ and FeZr$_2$. From analyses of lattice specific heat, we found that Ni substitution results in a systematic decrease in oscillator strength for the Einstein modes with 8.74 meV (CoZr$_2$). From phonon calculations, the low-energy optical phonon branches at the Γ point were observed for CoZr$_2$ and FeZr$_2$ with $c$-axis NTE, but not in NiZr$_2$ with positive thermal expansion. The enhancement of phonon density of states near the above-mentioned optical phonon energy in CoZr$_2$ and FeZr$_2$ is consistent with the specific heat analyses. We propose the importance of the low-energy optical phonons to the emergence of the $c$-axis NTE in $Tr$Zr$_2$.

## I. INTRODUCTION

Negative thermal expansion (NTE) materials that contract upon heating have the potential to tune the thermal expansion of materials. Control of thermal expansion property of materials is important for producing high-performance devices because even such a small expansion can degrade the performance of devices in modern advanced technology [1]. $ZrW_2O_8$ is a typical NTE material, and the striking feature is the wide temperature range (0.3 K < $T$ < 1050 K) NTE caused by the unique low-energy phonon mode [2-4]. The low-energy phonon modes propagate without distorting constituent units, $ZrO_6$ octahedra and $WO_4$ tetrahedra, which are known as Rigid Unit Modes (RUMs). The mechanism of NTE in $ZrW_2O_8$ is based on the RUMs which cause collective rotations of $ZrO_6$ octahedra and $WO_4$ tetrahedra [5]. Another typical NTE mechanism is related to electronic structure transition. For example, the manganese antiperovskite nitride shows sharp volume change because of the first-order antiferromagnetic transition, called as magnet volume effect [6,7]. In addition, in superconductors, NTE has been observed mostly below superconducting transition temperature ($T_c$), which is typically quite low. NTE phenomena in superconductors can be achieved by an evolution of the superconducting order parameter [8,9]. For NTE materials associated with electronic structure transition such as magnetic compounds and superconductors, realizing the wide-temperature range NTE is quite rare because the NTE can be realized by order-disorder phase transitions. If we could achieve volumetric zero thermal expansion with a wide temperature range in a superconductor, it is useful for superconductor devices or junctions with nano-scale structure or robustness to heat cycles between quite low temperature and room temperature.

Recently, we revealed that the $CoZr_2$ superconductor ($T_c$ ~ 5 K) shows uniaxial $c$-axis NTE in a wide temperature range (7 K < $T$ < 573 K) [10]. Note that the volume itself expands upon heating because the $a$-axis shows the usual PTE. According to the neutron scattering experiment, there are no crystal structural transitions or magnetic ordering down to 7 K. Therefore, the cause of uniaxial $c$-axis NTE should be not related to magnetic ordering, which is incontrast to the case of the manganese antiperovskite nitride. We can fully or partially substitute the Co site with other transition metals ($Tr$) in the tetragonal $CuAl_2$-type $TrZr_2$. We succeed in tuning the uniaxial $c$-axis NTE by chemical substitution of Co by Ni [11]. The tunable uniaxial $c$-axis NTE is highly correlated with a ratio of lattice parameters $c/a$, large $c/a$ ($c/a > 0.84$) favors leading the uniaxial $c$-axis NTE; for example, $Tr$ = Co, (Fe, Co, Ni), (Fe, Co, Rh, Ir), and so on [12]. However, small $c/a$ leads to PTE along the $c$-axis; therefore, the uniaxial $c$-axis NTE is absent for $Tr$ = Ni. $TrZr_2$ superconductors would play an important role in the implementation of modern nanoscale superconducting devices which are delicate to slight thermal expansion because of the tunable thermal expansion property. Superconducting devices containing Josephson junctions possibly can be applied to high-speed and energy-sufficient circuits [13]. Therefore, the development of $TrZr_2$ superconductors and elucidation of mechanisms causing the uniaxial $c$-axis NTE are necessary.

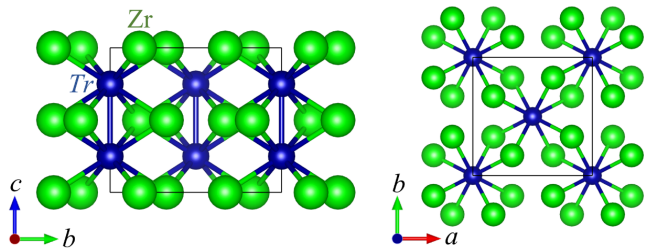

FIG. 1. Schematic images of the crystal structure of $CuAl_2$-type $TrZr_2$.

Figure 1 shows the crystal structure of tetragonal $CuAl_2$-type $TrZr_2$. According to density functional theory (DFT) calculations, the presence of relatively high-frequency optical phonon modes with huge negative Grüneisen parameters is the cause of the huge uniaxial $c$-axis NTE in $FeZr_2$ [14]. In $FeZr_2$, there are optical phonon modes with negative $\gamma_c$ (Grüneisen parameter along the $c$-axis) which generate soft Fe-Fe bonds along the $c$-axis. Moreover, the authors observed a large thermal expansion of Fe-Zr bonds through the EXAFS (extended X-ray absorption fine structure) experiments. Based on the soft Fe-Fe bonds and strong expansion of Fe-Zr bonds, Fe-Fe bond length can easily contract to maintain stable $Tr_2Zr_4$ octahedra, resulting in the uniaxial $c$-axis NTE. Although theoretical approaches have been shown, the experimental investigation of the optical phonon modes has not been performed for $Tr$Zr$_2$. Systematic analyses of phonons for solid solution systems are required to obtain more information on the cause of $c$-axis NTE in $Tr$Zr$_2$ and to discuss the importance of the optical phonons with negative $\gamma_c$. This article uses specific heat to analyze the experimental changes in optical phonons by $Tr$-site substitution. The experimental results are discussed in conjunction with the phonon calculation results.

## II. METHODS
### A. EXPERIMENTAL METHODS

Polycrystalline samples of FeZr$_2$ and Co$_{1-x}$Ni$_x$Zr$_2$ ($x$ = 0, 0.2, 0.4, 0.6, 0.8, 1) were synthesized through arc melting using Fe powder (Kojundo Chemical 99.9%), Co rods (Nilaco Corporation 99.98%), Ni wires (Nilaco Corporation 99%), and Zr plates (Nilaco Corporation 99.2%). Pelletized Fe powders and Co, Ni, and Zr solids were weighed as stoichiometric ratios. The atmospheric gas inside the arc furnace was replaced 3 times by pure Ar gas. Then, we started to melt the prepared pellet and the solids in the Ar atmosphere. A Ti ingot was used to absorb residual O$_2$ gas before synthesis. The samples were melted several times and turned over at each melting to improve homogeneity.

X-ray diffraction (XRD) patterns of the obtained polycrystalline samples were collected by $\theta$-$2\theta$ method with Cu-K$\alpha$ radiation on a Miniflex-600 (RIGAKU) diffractometer equipped with a high-resolution semiconductor detector D/tex-Ultra. The crystal structure of FeZr$_2$ and Co$_{1-x}$Ni$_x$Zr$_2$ ($x$ = 0, 0.2, 0.4, 0.6, 0.8, 1) were confirmed as CuAl$_2$-type tetragonal structure ($I4/mcm$) by refining the XRD patterns using RIETAN-FP [15]. The obtained XRD patterns are shown in Fig. S1(a) of the Supplemental Material. We confirmed that Ni was successfully substituted in Co$_{1-x}$Ni$_x$Zr$_2$ ($x$ = 0, 0.2, 0.4, 0.6, 0.8, 1) system because lattice constants $a$ and $c$ were systematically changed with increasing Ni amount (see Fig. S1(b) of the Supplemental Materials). We found that FeZr$_2$ contains impurity phases of Fe$_2$Zr and Zr, whereas Co$_{1-x}$Ni$_x$Zr$_2$ ($x$ = 0, 0.2, 0.4, 0.6, 0.8, 1) were confirmed as almost single-phase quality (see Figs. S1(c) and S1(d) of Supplemental Materials). The actual compositions of the obtained samples were consistent to nominal compositions (see Table S1 of Supplemental Materials), which were examined by energy-dispersive X-ray spectroscopy (EDX) method on a scanning electron microscope (SEM) TM-3030plus (Hitachi High-Tech) equipped with detector system SwiftED (Oxford). SEM images and elemental mapping images of FeZr$_2$ and Co$_{0.6}$Ni$_{0.4}$Zr$_2$ are shown in Figs. S2(a) and S2(b) of the Supplemental Materials, respectively. The schematic images of crystal stricture were depicted using VESTA [16].

Specific heat measurements were performed by thermal relaxation method using a Physical Property Measurement System (PPMS Dynacool, Quantum Design). The samples were fixed on a sample stage with the Apiezon N grease.

Sound velocity measurements were performed using 1077DATA (KARL DEUTSCH), and the longitudinal and transverse sound velocities were separately determined.

### B. CALCULATION METHODS

Our calculations have been carried out using density functional theory (DFT) as instated in the Vienna Ab initio Simulation Package (VASP) [17-19] and QUANTUM ESPRESSO package (QE) [20,21]. In VASP, the exchange-correlation energies were considered as described by employing the Perdew, Burke, and Ernzerhof (PBE) functional [22] within the generalized gradient approximation (GGA). Projector-augmented waves (PAWs) [23] were used to represent the electron-ion interactions. A 12×12×12 Monkhorst–Pack k-point [24] mesh was used for structural properties. The convergence criteria for energy and force were 1×10$^{-6}$ and 1×10$^{-4}$, respectively. The kinetic energy cutoff for plane wave expansion was set to 500 eV. This k-point and energy cutoff were sufficient for obtaining well-converged structural and electronic structure results.

For phonon dispersion curves, the PHONOPY code was used through density functional perturbation theory [25] employing a 4×4×4 supercell approximation. However, to determine the thermal expansion property, it is necessary to perform numerous phonon calculations for lattice parameters optimized at lower and higher values. Thermal expansion was predicted by the quasi-harmonic approximation (QHA). The most conventional approach for calculating thermal expansion based on QHA is directly minimizing the free energy. Here, we utilized machine learning capabilities within VASP to address the high cost associated with these extensive phonon calculations. This facilitated the seamless integration of phonon dispersion analysis into our research workflow. This method highlights the effective combination of machine learning and on-the-fly phonon calculations, offering a robust approach for studying lattice vibrations and enhancing our comprehension of material properties crucial for diverse applications.

The superconductivity properties were also studied utilizing the Quantum ESPRESSO (QE) code [20,21]. For exchange–correlation functional Perdew-Burke-Ernzerhof (PBE) type [22], Generalized Gradient approximations (GGA) and ultrasoft pseudopotentials with scalar relativistic were used. Broyden–Fletcher–Goldfarb–Shanno (BFGS) algorithm [26] was chosen to optimize geometry. The Methfessel–Paxton broadening technique [27] was used with the smearing parameter 0.03 Ry. The kinetic energy cut-offs for wave functions and charge density were set to 60 Ry and 480 Ry, respectively. For Brillouin zone integration with a 16×16×16 k-point grid for self-consistent calculations. The electron-phonon matrix elements were computed using linear response theory [20,21] and the Migdal-Eliashberg theory [28,29]. McMillan's equation, modified by Allen and Dynes [30-32] was applied to estimate the superconducting transition temperature ($T_c$) as follows:

$$T_c = \frac{\omega_{\log}}{1.20} \exp\left[-\frac{1.04(1+\lambda)}{\lambda - \mu^*(1+0.62\lambda)}\right],$$

where the parameter $\mu^*$ indicates effective Coulomb repulsion, with values usually ranging between 0.10 and 0.16. The logarithmically averaged frequency $\omega_{\log}$ was defined as

$$\ln\omega_{\log} = \frac{2}{\lambda} \int d\omega \frac{\alpha^2 F(\omega)}{\omega} \ln\omega,$$

where $\alpha^2 F(\omega)$ is the Eliashberg phonon spectral function. The electron-phonon coupling constant $\lambda$ is calculated by:

$$\lambda = 2\int \frac{\alpha^2 F(\omega)}{\omega} d\omega.$$

## III. RESULTS

First, we analyze the specific heat data of $CoZr_2$, $FeZr_2$, and $NiZr_2$. The temperature dependences of the total specific heat $C(T)$ under zero field are shown in Fig. 2(a). We confirmed that values of $C(T)$ at around 300 K are close to the Dulong-Petit law value, $3nR = 74.8$ J/K mol, where $n = 3$ is the number of atoms per formula unit, and $R \approx 8.31$ J/K mol is an ideal gas constant. Figure 2(b) shows the squared-temperature dependences of the total specific heat under zero field in the form of $C(T)/T$. $CoZr_2$ shows a superconducting transition at around 5 K under zero field, which is consistent with previous works on $CoZr_2$ [33,34]. No bulk superconducting transition was detected by specific heat for $FeZr_2$ and $NiZr_2$ under zero field above 1.8 K. The $C(T)/T$ data between 40 $K^2$ < $T^2$ < 100 $K^2$ were fitted to $C(T)/T = \gamma + \beta T^2$, where $\gamma$ and $\beta$ are Sommerfeld coefficient and coefficient of phonon contribution to total specific heat, respectively. We can extract only phonon contribution to total specific heat $C_{lat}(T)$ by subtracting electronic specific heat $\gamma T$ from $C(T)$. Figure 2(c) shows the temperature dependences of phonon contribution to total specific heat in the form of $C_{lat}(T)/T^3$ between 7 K < $T$ < 300 K for $CoZr_2$, $FeZr_2$, and $NiZr_2$. We found a peak structure (Schottky-type specific heat anomaly) in $C_{lat}(T)/T^3$ at around 20 K for $CoZr_2$ and $FeZr_2$ whereas the peak structure is absent in $NiZr_2$. The observed peak structure in $C_{lat}(T)/T^3$ for $CoZr_2$ and $FeZr_2$ would be related to the optical phonon modes which were proposed to be a possible cause of the uniaxial $c$-axis NTE in $FeZr_2$ [14]. DFT calculations revealed that giant $c$-axis NTE observed in $FeZr_2$ is mainly driven by the optical phonon branches with relatively high energy above roughly 8 meV [14]. In $FeZr_2$, the high-frequency optical phonon modes excite multiple, large, and negative $\gamma_c$; therefore, the uniaxial $c$-axis NTE is induced. $NiZr_2$ exhibits uniaxial $c$-axis PTE because the optical phonon branches with negative $\gamma_c$ are absent. Therefore, we expect that the optical phonon branches with negative $\gamma_c$ are present for $CoZr_2$ as in $FeZr_2$ and contribute to the emergence of the uniaxial $c$-axis NTE in a wide temperature range [10].

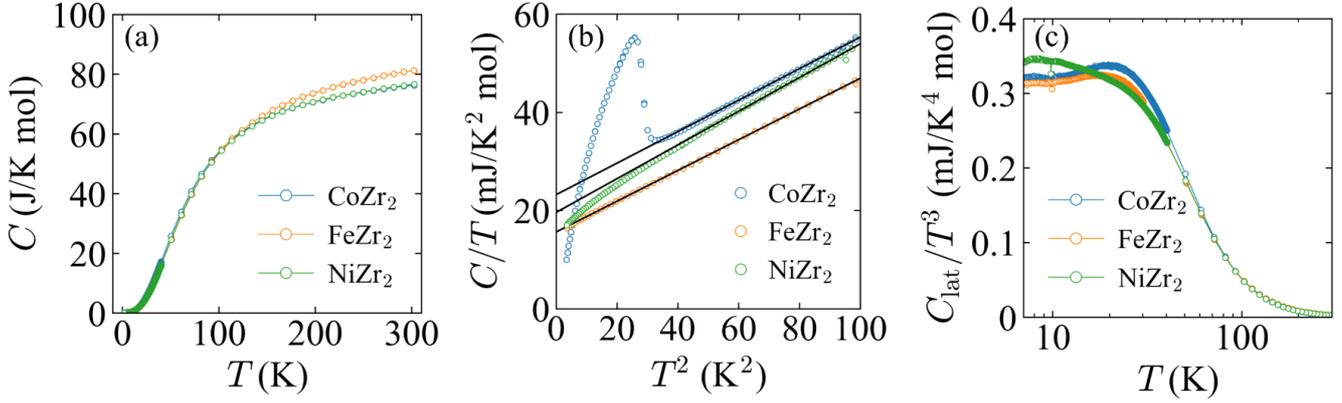

FIG. 2 (a) Temperature dependences of total specific heat $C(T)$ under zero field for CoZr$_2$, FeZr$_2$, and NiZr$_2$. (b) Squared temperature dependences of total specific heat under zero field in the form of $C(T)/T$. Solid lines are the fit of $C(T)/T = \gamma + \beta T^2$. (c) Temperature dependences of phonon contribution to total specific heat in the form of $C_{lat}(T)/T^3$ for CoZr$_2$, FeZr$_2$, and NiZr$_2$.

To discuss the optical phonon contributions to the uniaxial $c$-axis NTE, we analyze the $C_{lat}(T)/T^3$ data using a model based on the combination with the Debye and Einstein models. Related analyses have been attempted in phonon-driven flexible network NTE materials for instance ZrW$_2$O$_8$ [3]. From neutron scattering experiments, the low-energy phonon modes with 1.5–8.5 meV are suggested to play an important role in NTE [4] for ZrW$_2$O$_8$, and the low-energy phonon modes were quantitively analyzed by the Einstein model reflecting optical phonon branches [3]. In our analysis for CoZr$_2$, FeZr$_2$, and NiZr$_2$, we adopt a model with a single Debye model and double Einstein models to obtain quantitative insight into optical phonon related to $c$-axis thermal expansion. Debye model $C_D(T)$ and Einstein model $C_E(T)$ per atom in J/K mol unit are expressed in the following equations.

$$C_D(T) = 9R\left(\frac{T}{\Theta_D}\right)^3 \int_0^{\frac{\Theta_D}{T}} \frac{x^4 e^x}{(e^x - 1)^2} dx,$$

$$C_E(T) = 3R\left(\frac{\Theta_E}{T}\right)^2 \frac{\exp\left(\frac{\Theta_E}{T}\right)}{\left[\exp\left(\frac{\Theta_E}{T}\right) - 1\right]^2},$$

where $\Theta_D$ and $\Theta_E$ are Debye temperature and Einstein temperature, respectively. Note that this combination model tries to explain the total phonon density of states (DOS) by a single Debye function, but the Schottky-type specific heat contribution is fitted using the Einstein functions. The number of atoms in the formula unit is 3 for CoZr$_2$, FeZr$_2$, and NiZr$_2$; therefore, in our model, phonon contribution to total specific heat is described as $C_{lat}(T) = n_D C_D(T) + n_{E1} C_E(T) + n_{E2} C_E(T)$ with the summation rule of $n_D + n_{E1} + n_{E2} = 3$ where $n$ is the oscillator strength that is the parameter of the weight of each model related to the assumed partial DOS. We determined the parameters of oscillator strength ($n_D$, $n_{E1}$, and $n_{E2}$), $\Theta_D$, and $\Theta_E$ as most explain the $C_{lat}(T)/T^3$ data. The fitting results of $C_{lat}(T)/T^3$ for CoZr$_2$, FeZr$_2$, and NiZr$_2$ are shown in Figs. 3(a), 3(b), and 3(c), respectively. The fitting parameters are summarized in TABLE I. The $\Theta_D$ values were estimated through the Debye modes as 274 K ($n_D$ = 2.8898) for CoZr$_2$, 277 K ($n_D$ = 2.9019) for FeZr$_2$, and 280 K ($n_D$ = 2.9331) for NiZr$_2$. We can consider the peak structures of $C_{lat}(T)/T^3$ observed in CoZr$_2$ and FeZr$_2$ are responsible for E1 modes with $\hbar\omega_{E1}$ = 8.74 meV ($n_{E1}$ = 0.1068) and $\hbar\omega_{E1}$ = 8.90 meV ($n_{E1}$ = 0.0931), respectively. The fitting revealed that the estimated E1 mode energy of FeZr$_2$ corresponds to the optical phonon energy which plays an important role in emerging uniaxial $c$-axis NTE suggested by DFT calculations [14]. We also confirmed that the E1 mode energy and oscillator strength of CoZr$_2$ are comparable with those of FeZr$_2$. Therefore, the uniaxial $c$-axis NTE in CoZr$_2$ and FeZr$_2$ would be attributed to the optical phonon modes corresponding to the E1 modes. For analysis of NiZr$_2$, we found E1 mode energy $\hbar\omega_{E1}$ = 7.67 meV is close to that of CoZr$_2$ and FeZr$_2$; however, $n_{E1}$ = 0.0598 is clearly smaller than that of CoZr$_2$ and FeZr$_2$. Therefore, the phonon DOS of the E1 mode for NiZr$_2$ is not dominant to the $C_{lat}(T)$, and the fitting result is consistent with the absence of optical phonon branches with negative $\gamma_c$ leading the uniaxial *c*-axis NTE [14]. The Einstein mode with lower energy (E2) is considered to hardly contribute to the emergence of NTE for CoZr$_2$, FeZr$_2$, and NiZr$_2$ because the oscillator strength is lowest for others.

Furthermore, we analyze specific heat data of the Co$_{1-x}$Ni$_x$Zr$_2$ ($x$ = 0, 0.2, 0.4, 0.6, 0.8, 1) solid-solution system and evaluate optical phonon contributions. We recently succeeded in controlling (switching) thermal expansion along the *c*-axis by Ni substitution for Co site in CoZr$_2$. With increasing Ni amount, the uniaxial *c*-axis NTE gradually changes to uniaxial *c*-axis PTE [11]. Based on the success of controlling the thermal expansion along the *c*-axis by chemical substitution, we expect that the peak structure of $C_{\text{lat}}(T)/T^3$ observed in CoZr$_2$ gradually disappears with increasing Ni amount. Figure 4 shows the fitting results of $C_{\text{lat}}(T)/T^3$ for Co$_{1-x}$Ni$_x$Zr$_2$ ($x$ = 0, 0.2, 0.4, 0.6, 0.8, 1) using the similar model with single Debye model and double Einstein models. The fitting parameters are summarized in TABLE II. The clear peak structures of $C_{\text{lat}}(T)$ were observed in CoZr$_2$ ($x$ = 0) and Co$_{0.8}$Ni$_{0.2}$Zr$_2$ ($x$ = 0.2) whereas the peak structures were absent in $x$ = 0.4, 0.6, 0.8, and 1. The observation results of $C_{\text{lat}}(T)/T^3$ are consistent with the change of uniaxial *c*-axis thermal expansion from NTE to PTE by Ni substitution because the uniaxial *c*-axis NTE region of Co$_{1-x}$Ni$_x$Zr$_2$ system mainly is $x$ = 0–0.3 [11]. The oscillator strength of the E1 mode ($n_{\text{E1}}$) gradually decreased with increasing Ni amount, which indicates the contribution of the E1 mode related to the uniaxial *c*-axis NTE to $C_{\text{lat}}(T)$ becomes smaller. In the discussion part, we show phonon calculation results and discuss the possible mechanisms of the emergence of the *c*-axis NTE in *Tr*Zr$_2$.

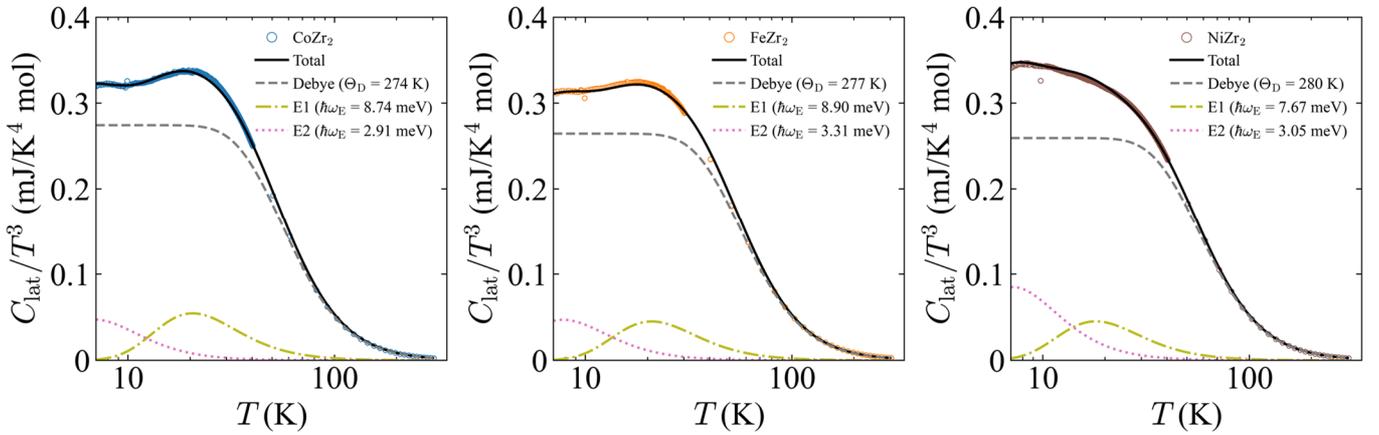

FIG. 3. Fitting results of $C_{\text{lat}}(T)/T^3$ using a single Debye model and double Einstein models (E1 and E2) for (a) CoZr$_2$, (b) FeZr$_2$, and (c) NiZr$_2$.

TABLE I. Fitting parameters of $C_{\text{lat}}(T)/T^3$ for CoZr$_2$, FeZr$_2$, and NiZr$_2$.

| Mode | Energy | Oscillator strength |
|---|---|---|
| **CoZr$_2$** | | |
| Debye | 274 K | 2.8898 |
| E1 | 8.74 meV | 0.1068 |
| E2 | 2.91 meV | 0.0034 |
| **FeZr$_2$** | | |
| Debye | 277 K | 2.9019 |
| E1 | 8.90 meV | 0.0931 |
| E2 | 3.31 meV | 0.0050 |
| **NiZr$_2$** | | |
| Debye | 280 K | 2.9331 |
| E1 | 7.67 meV | 0.0598 |
| E2 | 3.05 meV | 0.0071 |

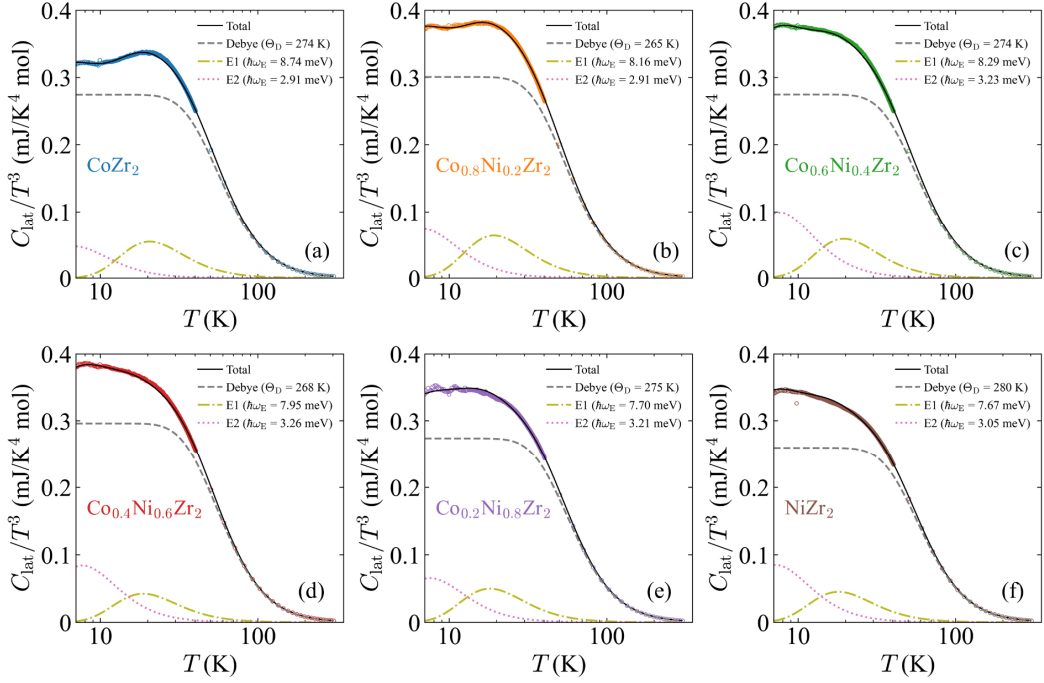

FIG. 4. Fitting results of $C_{lat}(T)/T^3$ using a single Debye model and double Einstein models (E1 and E2) for (a) CoZr$_2$ ($x = 0$), (b) Co$_{0.8}$Ni$_{0.2}$Zr$_2$ ($x = 0.2$), (c) Co$_{0.6}$Ni$_{0.4}$Zr$_2$ ($x = 0.4$), (d) Co$_{0.4}$Ni$_{0.6}$Zr$_2$ ($x = 0.6$), (e) Co$_{0.2}$Ni$_{0.8}$Zr$_2$ ($x = 0.8$), and (f) NiZr$_2$ ($x = 1$)

TABLE II. Fitting parameters of $C_{lat}(T)/T^3$ for Co$_{1-x}$Ni$_x$Zr$_2$ ($x = 0, 0.2, 0.4, 0.6, 0.8$, and 1).

| | CoZr$_2$ ($x = 0$) | | | Co$_{0.8}$Ni$_{0.2}$Zr$_2$ ($x = 0.2$) | |
|---|---|---|---|---|---|
| Mode | Energy | Oscillator strength | Mode | Energy | Oscillator strength |
| Debye | 274 K | 2.8898 | Debye | 265 K | 2.8932 |
| E1 | 8.74 meV | 0.1068 | E1 | 8.16 meV | 0.1015 |
| E2 | 2.91 meV | 0.0034 | E2 | 2.91 meV | 0.0053 |
| | Co$_{0.6}$Ni$_{0.4}$Zr$_2$ ($x = 0.4$) | | | Co$_{0.4}$Ni$_{0.6}$Zr$_2$ ($x = 0.6$) | |
| Mode | Energy | Oscillator strength | Mode | Energy | Oscillator strength |
| Debye | 274 K | 2.8921 | Debye | 268 K | 2.92896 |
| E1 | 8.29 meV | 0.0981 | E1 | 7.95 meV | 0.06248 |
| E2 | 3.23 meV | 0.0098 | E2 | 3.26 meV | 0.00856 |
| | Co$_{0.2}$Ni$_{0.8}$Zr$_2$ ($x = 0.8$) | | | NiZr$_2$ ($x = 1$) | |
| Mode | Energy | Oscillator strength | Mode | Energy | Oscillator strength |
| Debye | 275 K | 2.92682 | Debye | 280 K | 2.9331 |
| E1 | 7.70 meV | 0.06680 | E1 | 7.67 meV | 0.0598 |
| E2 | 3.21 meV | 0.00638 | E2 | 3.05 meV | 0.0071 |

In addition, we measured transverse ($v_t$) and longitudinal ($v_l$) sound velocities for CoZr$_2$ and NiZr$_2$. The sound velocities at room temperature were $v_t$ = 1736 m s$^{-1}$ (1801 m s$^{-1}$) and $v_l$ = 2968 m s$^{-1}$ (4300 m s$^{-1}$) for CoZr$_2$ (NiZr$_2$). An average sound velocity $v_s$ can be calculated using the following equation:

$$v_s = \left[\frac{1}{3}\left(\frac{2}{v_t^3} + \frac{1}{v_l^3}\right)\right]^{-\frac{1}{3}}.$$

We obtained $v_s$ = 1925 m s$^{-1}$ and 2037 m s$^{-1}$ for CoZr$_2$ and NiZr$_2$, respectively. From obtained $v_s$ values, we can calculate $\Theta_D$ values of CoZr$_2$ and NiZr$_2$ by the following equation:

$$\Theta_D = \frac{\hbar}{k_B}\left(\frac{6\pi^2 N}{V}\right)^{\frac{1}{3}} v_s,$$

where $\hbar \approx 1.05\times10^{-34}$ J s is a reduced Planck constant, $k_B \approx 1.38\times10^{-23}$ J K$^{-1}$ is a Boltzmann constant, and $N = 12$ is the number of atoms in the unit cell. $V \approx 226.98$ Å$^3$ (224.87 Å$^3$) is a calculated volume of the unit cell by Rietveld refinement for CoZr$_2$ (NiZr$_2$). The calculated $\Theta_D$ values were 215 K and 228 K for CoZr$_2$ and NiZr$_2$, respectively; the trend agrees with the estimated $\Theta_D$ values from the specific heat data analysis.

## IV. DISCUSSION

From the specific heat analyses, we revealed that the optical-phonon (~8.7 meV) contributions to $C_{lat}$ in CoZr$_2$ are weakened by Ni substitution, and the uniaxial $c$-axis NTE is suppressed and switches to uniaxial $c$-axis PTE by Ni substitution in Co$_{1-x}$Ni$_x$Zr$_2$ [11]. Here, we discuss the experimental results by comparing them with phonon calculations of CoZr$_2$, FeZr$_2$, and NiZr$_2$. The connection implies that conducting phonon calculations is essential in understanding the effects of introducing Ni into the Co$_{1-x}$Ni$_x$Zr$_2$ compound. The phonon dispersion within CoZr$_2$, FeZr$_2$, and NiZr$_2$, each characterized by a primitive cell comprising six atoms and a total of eighteen phonon branches, consisting of three acoustic and fifteen optical branches. Through computations across high symmetry points in the Brillouin zone (as depicted in Fig. 5 (a) for CoZr$_2$, Fig. 6 (a) for FeZr$_2$, and Fig. 7(a) for NiZr$_2$), we confirm the compounds' dynamical stability, observing an absence of negative frequencies. The crossing between acoustic and optical modes is notably observed at the Γ—X—M—Γ—Z—P—N—Z1—M high symmetry points within the Brillouin zone for all three compounds. While the highest frequencies of the optical mode vary, with CoZr$_2$ and NiZr$_2$ surpassing 6 THz and FeZr$_2$ exceeding 7 THz, the lowest optical phonon branches are consistently found at the Γ point, with the highest at different symmetry points. This results in the appearance of clear peaks in phonon density of states (PDOS), and the larger optical PDOS in CoZr$_2$ and FeZr$_2$ than in NiZr$_2$ is consistent with the specific heat analyses. Additionally, the highest acoustic modes vary in location between the Γ and Z points. Analysis of the total and partial phonon density of states across these materials, depicted in respective figures (Fig. 5 (b) for CoZr$_2$, Fig. 6 (b) for FeZr$_2$, and Fig. 7 (b) for NiZr$_2$), reveals distinct contributions, with Zr primarily contributing to low frequencies, while Co dominates in CoZr$_2$ and Fe alongside Zr in FeZr$_2$ at higher frequencies due to their significant atomic mass differences. These findings shed light on the diverse phonon behaviors in these compounds, providing valuable insights into their dynamic properties and potential applications in various fields. The volume-temperature relationships and thermal expansion behaviors for CoZr$_2$ are illustrated in Figs. 5(c-d), while for FeZr$_2$ and NiZr$_2$, these can be found in Figs. 6(c-d) and Figs. 7(c-d), respectively. Concerning the variation of volume with temperature, an intriguing trend emerges. In the case of CoZr$_2$, we note a negative slope of thermal expansion between 250 K and 500 K, indicating the suppression of thermal expansion as temperature increases. This trend is mirrored in FeZr$_2$, suggesting a similar behavior. However, NiZr$_2$ exhibits larger thermal expansion coefficients and a monotonous increase in thermal expansion coefficient, consistent with the presence of normal PTE in NiZr$_2$. These experimental findings align well with our expectations and contribute to our understanding of the distinct thermal expansion properties exhibited by these compounds.

Finally, we calculate the $T_c$ of the materials using the calculation results. Figures 5(e), 6(e), and 7(e) show a detailed analysis of $\alpha^2 F(\omega)$ and the integrated electron-phonon coupling constant $\lambda$ across different frequencies, accounting for an effective Coulomb repulsion of $\mu^* = 0.16$. This analysis continues in Table III, where we record the logarithmically averaged frequency $\omega_{log}$, the electron-phonon coupling constant, and the resulting $T_c$. Remarkably, our computed $T_c$ values for these materials exhibit a good agreement with experimental observations, emphasizing the strength and precision of our theoretical framework in revealing the complex mechanisms that govern superconductivity in these materials. An analysis of the $\Theta_D$ for CoZr$_2$, FeZr$_2$, and NiZr$_2$ was conducted, providing theoretical values of 232 K, 207 K, and 267 K, respectively. The trend that the highest $\Theta_D$ is observed in NiZr$_2$ is consistent with experimental $\Theta_D$ values shown above.

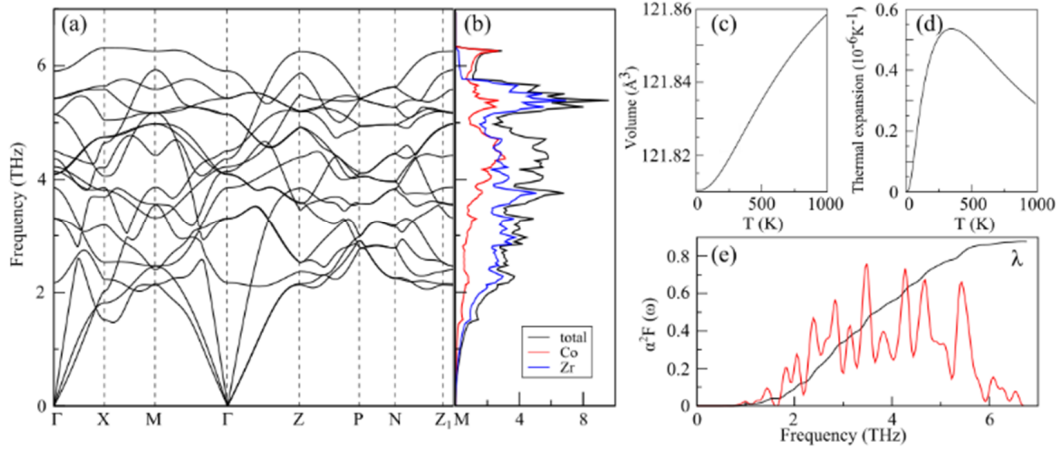

FIG. 5. The calculated (a) phonon dispersion curves, (b) total and partial phonon density of states, (c) Volume-temperature and (d) thermal expansion coefficient-temperature relations and (e) Eliashberg spectral function $F(\omega)$ and integrated electron-phonon coupling constant $\lambda$ of $CoZr_2$.

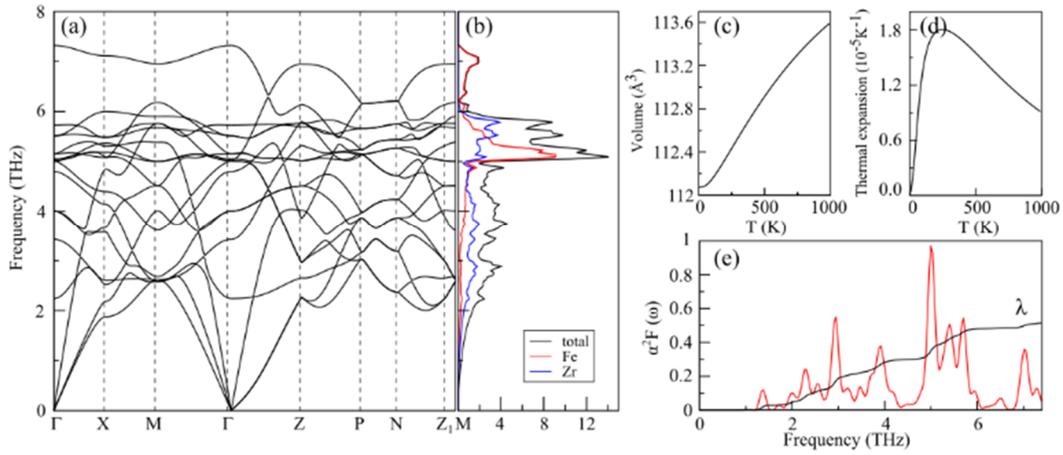

FIG. 6. The calculated (a) phonon dispersion curves, (b) total and partial phonon density of states, (c) Volume-temperature and (d) thermal expansion coefficient-temperature relations and (e) Eliashberg spectral function $F(\omega)$ and integrated electron-phonon coupling constant $\lambda$ of $FeZr_2$.

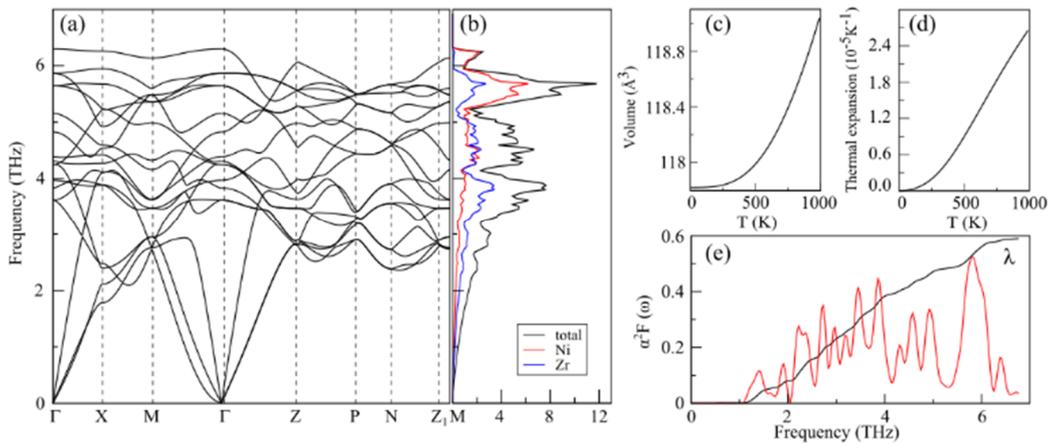

FIG. 7. The calculated (a) phonon dispersion curves, (b) total and partial phonon density of states, (c) Volume-temperature and (d) thermal expansion coefficient-temperature relations and (e) Eliashberg spectral function $F(\omega)$ and integrated electron-phonon coupling constant $\lambda$ of $NiZr_2$.

TABLE III. Computed $\lambda$, $T_c$ and $\Theta_D$ values for CoZr$_2$, FeZr$_2$, and NiZr$_2$. For $T_c$ calculation, $\mu^* = 0.16$ is taken. Experimental $\Theta_D$ values are taken from specific heat analyses.

| Compounds | $\lambda$ | $T_c$ (K) Calc. | $T_c$ (K) Expt. | $\Theta_D$ (K) Calc. | $\Theta_D$ (K) Expt. |
|---|---|---|---|---|---|
| CoZr$_2$ | 0.87 | 5.87 | 5.5 [35], 5.0 [36], 6.3 [37] | 232 | 274 |
| FeZr$_2$ | 0.51 | 0.81 | 0.17 [35,38] | 207 | 277 |
| NiZr$_2$ | 0.59 | 1.56 | 1.6, 1.52 [35,37] | 267 | 280 |

## V. SUMMARY

The physical properties of the polycrystalline samples of CuAl$_2$-type $Tr$Zr$_2$ have been studied experimentally (specific heat and sound velocity experiments) and theoretically (phonon calculations) to discuss the possible causes of $c$-axis NTE in CoZr$_2$ and FeZr$_2$. From analyses of $C_{lat}$, we found that Ni substitution results in a systematic decrease in oscillator strength for the Einstein modes with 8.74 meV (CoZr$_2$). From phonon calculations, the low-energy optical phonon branches at the $\Gamma$ point were observed for CoZr$_2$ and FeZr$_2$ with $c$-axis NTE, but not in NiZr$_2$ with PTE. The enhancement of PDOS near the optical phonon energy in CoZr$_2$ and FeZr$_2$ is consistent with the specific heat analyses. We propose the importance of the low-energy optical phonons to the emergence of $c$-axis NTE in $Tr$Zr$_2$. The validity of the phonon calculations was confirmed by calculating the $T_c$ of the materials.

**Acknowledgments**

The authors thank O. Miura and T. Shiga for supports in experiments and discussion. This work is partly supported by JSPS-KAKENHI (No.: 23KK0088), TMU Research Project for Emergent Future Society, and Collaborative Research Project of Laboratory for Materials and Structures, Institute of Innovative Research, Tokyo Institute of Technology. This study was computationally supported by T.C. Strategy and Budget Directorate under Project No: 2019K12-92587.

# Supplemental Materials

**Specific heat analyses on optical-phonon-derived uniaxial negative thermal expansion system $Tr$Zr$_2$ ($Tr$ = Fe and Co$_{1-x}$Ni$_x$)**

Yuto Watanabe[1], Ceren Tayran[2,3]*, Md. Riad Kasem[1], Aichi Yamashita[1], Mehmet Çakmak[4,5], Takayoshi Katase[6], Yoshikazu Mizuguchi[1]*

[1]Department of Physics, Tokyo Metropolitan University, Hachioji, Tokyo 192-0397, Japan
[2]Department of Physics, Faculty of Science, Gazi University, 06500, Ankara, Turkey
[3]DIFFER–Dutch Institute for Fundamental Energy Research, De Zaale 20, 5612 AJ Eindhoven, The Netherlands
[4]Department of Photonics, Faculty of Applied Sciences, Gazi University, 06500, Ankara, Turkey
[5]Photonics Application and Research Center, Gazi University, 06500, Ankara, Turkey
[6]MDX Research Center for Element Strategy, International Research Frontiers Initiative, Tokyo Institute of Technology, Yokohama 226-8501, Japan

Corresponding authors: C. Tayran (c.tayran@gazi.edu.tr), Y. Mizuguchi (mizugu@tmu.ac.jp)

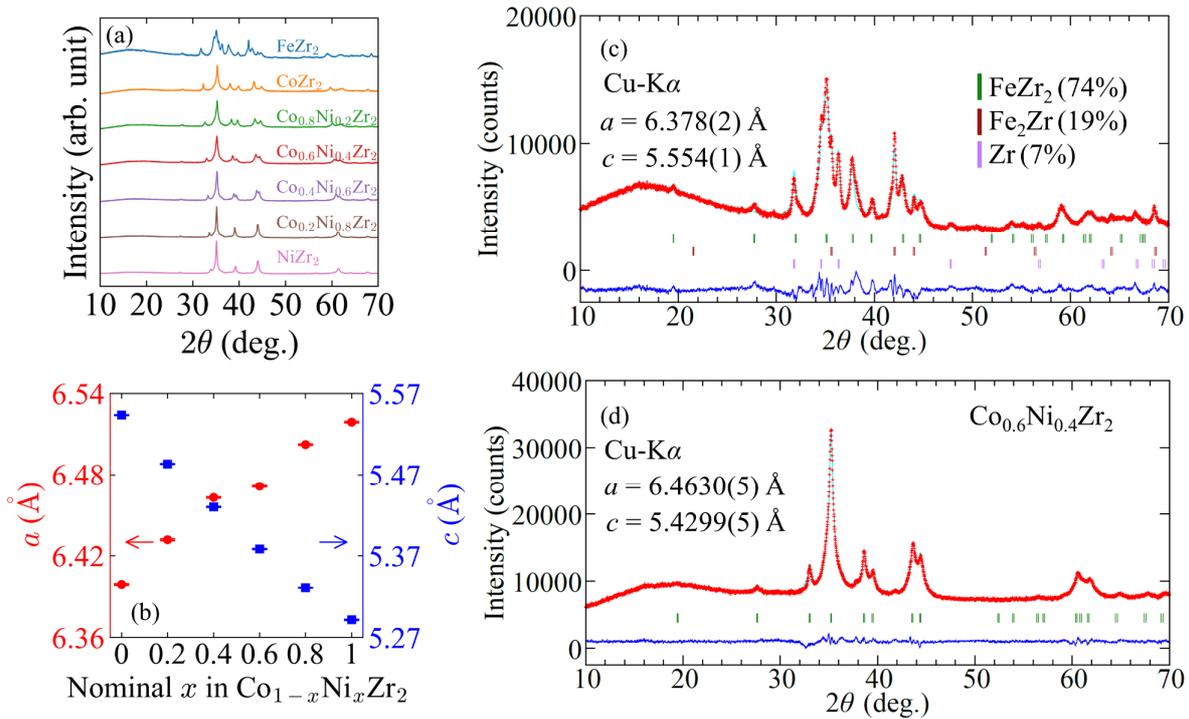

FIG. S1 (a) XRD patterns of polycrystalline FeZr$_2$ and Co$_{1-x}$Ni$_x$Zr$_2$ ($x$ = 0, 0.2, 0.4, 0.6, 0.8, 1) with Cu-K$\alpha$ radiation. (b) Nominal Ni amount ($x$) dependences of lattice constants $a$ and $c$ of Co$_{1-x}$Ni$_x$Zr$_2$. (c-d) Rietveld refinement results of FeZr$_2$ and Co$_{0.6}$Ni$_{0.4}$Zr$_2$, respectively. Impurity phase amounts of FeZr$_2$ are in mass fraction.

Table S1. Actual compositions of FeZr$_2$ and Co$_{1-x}$Ni$_x$Zr$_2$ ($x$ = 0, 0.2, 0.4, 0.6, 0.8, 1) obtained from EDX.

| $Tr$ | Co | Ni | Fe | Zr |
|---|---|---|---|---|
| Fe | - | - | 0.80(3) | 2.20(3) |
| Co | 0.94(5) | - | - | 2.06(5) |
| Co$_{0.8}$Ni$_{0.2}$ | 0.75(2) | 0.174(6) | - | 2.07(2) |
| Co$_{0.6}$Ni$_{0.4}$ | 0.537(3) | 0.358(4) | - | 2.105(4) |
| Co$_{0.4}$Ni$_{0.6}$ | 0.357(6) | 0.52(1) | - | 2.12(1) |
| Co$_{0.2}$Ni$_{0.8}$ | 0.165(3) | 0.653(7) | - | 2.182(5) |
| Ni | - | 0.87(1) | - | 2.13(1) |

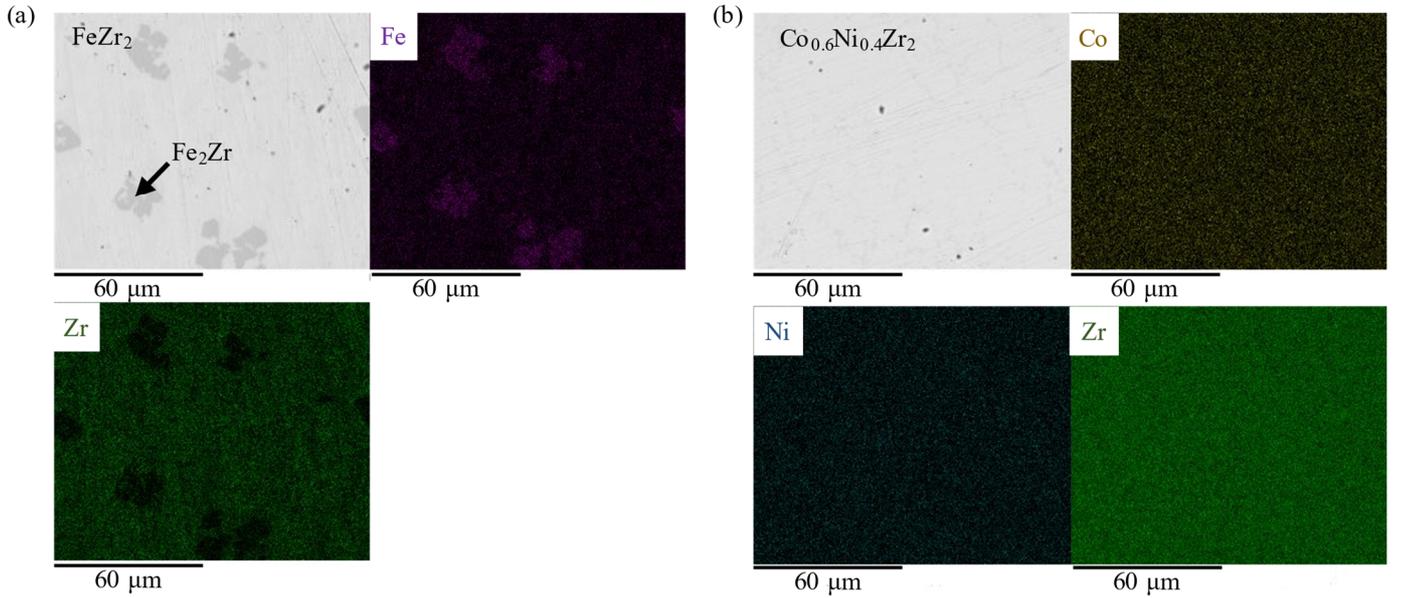

FIG. S2 (a-b) SEM images with back scattering electron and elemental mapping images for FeZr$_2$ and Co$_{0.6}$Ni$_{0.4}$Zr$_2$, respectively. For FeZr$_2$, we found a Fe-rich secondary phase (Fe$_2$Zr) denoted as a solid arrow.